\documentclass[aps,prb,twocolumn,floatfix,footinbib,showpacs,superscriptaddress]{revtex4-1}
\usepackage{graphicx}
\usepackage{amsfonts,amsmath,amssymb}
\usepackage{amsthm}
\usepackage{dsfont,bm}
\usepackage{color}
\usepackage{soul} 
\usepackage{amsbsy}
\usepackage[colorlinks=true,linkcolor=blue,pagecolor=blue,filecolor=blue,menucolor=blue,urlcolor=blue,citecolor=blue,anchorcolor=blue]{hyperref}%
\usepackage{sidecap}
\usepackage{txfonts}
\usepackage{amsbsy}
\usepackage{times} 
\usepackage{dcolumn}

\newcommand{\ua}{\uparrow}
\newcommand{\da}{\downarrow}

\newcommand{\rr}{({\bm r})}

\usepackage{amssymb,amsmath,bm}

\begin{document}

\title{Fraunhofer response and supercurrent spin switching in Black Phosphorus with strain and disorder}

\author{Mohammad Alidoust}
\affiliation{Department of Physics, K.N. Toosi University of Technology, Tehran 15875-4416, Iran}
\author{Morten Willatzen}
\affiliation{Beijing Institute of Nanoenergy and Nanosystems, Chinese Academy of Sciences, No. 30 Xueyuan Road, Haidian District, Beijing 100083, China}
\affiliation{Department of Photonics Engineering, Technical University of Denmark, DK-2800 Kongens Lyngby, Denmark}
\author{Antti-Pekka Jauho}
\affiliation{Center for Nanostructured Graphene (CNG), Department of Micro- and Nanotechnology, Technical University of Denmark, DK-2800 Kongens Lyngby, Denmark}

\date{\today}

\begin{abstract}
We develop theory models for both ballistic and disordered superconducting monolayer black phosphorus devices in the presence of magnetic exchange field and strain. The ballistic case is studied through a microscopic Bogoliubov-de Gennes formalism while for the disordered case we formulate a quasiclassical model. Utilizing the two models, we theoretically study the response of supercurrent to an externally applied magnetic field in two-dimensional black phosphorus Josephson junctions. Our results demonstrate that the response of the supercurrent to a perpendicular magnetic field in ballistic samples can deviate from the standard Fraunhofer interference pattern when the Fermi level and mechanical strain are varied. This finding suggests the combination of chemical potential and strain is an efficient external knob to control the current response in highly sensitive strain-effect transistors and superconducting quantum interference devices. We also study the supercurrent in a superconductor-ferromagnet-ferromagnet-superconductor junction where the magnetizations of the two adjacent magnetized regions are uniform with misaligned orientations. We show that the magnetization misalignment can control the excitation of harmonics higher than the first harmonic $\sin \varphi$ (in which $\varphi$ is the phase difference between the superconductors) in supercurrent and constitutes a full spin switching current element. Finally, we discuss possible experimental implementations of our findings. We foresee our models and discussions could provide guidelines to experimentalists in designing devices and future investigations.        
\end{abstract}
\maketitle

\section{introduction}\label{sec:intro}

Orthorhombic bulk black phosphorus (BP) under pressure experiences topological phase transitions, making phosphorus allotropes an attractive new research area\cite{Xiang,Guo,Zhao,Gong}. Among phosphorus allotropes, black phosphorus is the most stable crystal structure \cite{Bridgman,Keyes}. Possessing a direct bandgap that is tunable from $0.3$~eV to $2.0$~eV by applying strain; and electric field, manipulating the number of layers involved, and \textit{in situ} doping with different atoms such as $\rm K$ and $\rm Rb$\cite{kim1,kim2}, BP offers an excellent platform with great control over the density of charge carriers by different means \cite{Carvalho,wylsuperc_exp1,Rudenko,Pereira,wei,Y.Ren1,Hedayati}. These features of BP are applicable to atomically thin electronics, modern two-dimensional mechanical transistors, ultrasensitive sensors with high on-off ratios, and optomechanic devices working under blue or ultraviolet light \cite{Li1,Wang2,Wu1,Sangwan,liu1}. 

Recent research on phosphorus allotropes led to the experimental realization of phosphorene\cite{liu1}. Also, it has been found that black phosphorus in both orthorhombic and rhombohedral structures can intrinsically develop superconductivity \cite{Kawamura1,Karuzawa2,Flores,Wang1,Shirotani,Livas,Q.Huang,Jun-JieZhang,YanqingFeng,R.Zhang,Yuan}. However, the dominant pairing type in BP has brought up discussions and is still unclear. The same issue arises in even monolayer graphene which is a much simpler two-dimensional crystal. Particularly, several different and exotic pairing types have been proposed, ranging from the usual $s$-wave to $p+ip$, $d+id$, and $f$-wave pairings \cite{Ma,Faye,Nandkishore,Kiesel,Uchoa} and there is still no agreement about the dominant pairing symmetry. In effect, theoretical predictions of pairing symmetries in these systems strongly depend on the strength of interactions in the model considered, pairing mechanisms, and approximations made. Therefore, detailed experimental information on the predominant interactions and pairing mechanism is critically important to identify the pairing(s) that is (are) the most energetically favorable. 

\begin{SCfigure*}
\includegraphics[clip, trim=0.4cm 1.8cm 0.cm 1.0cm, width=11.0cm,height=6.50cm]{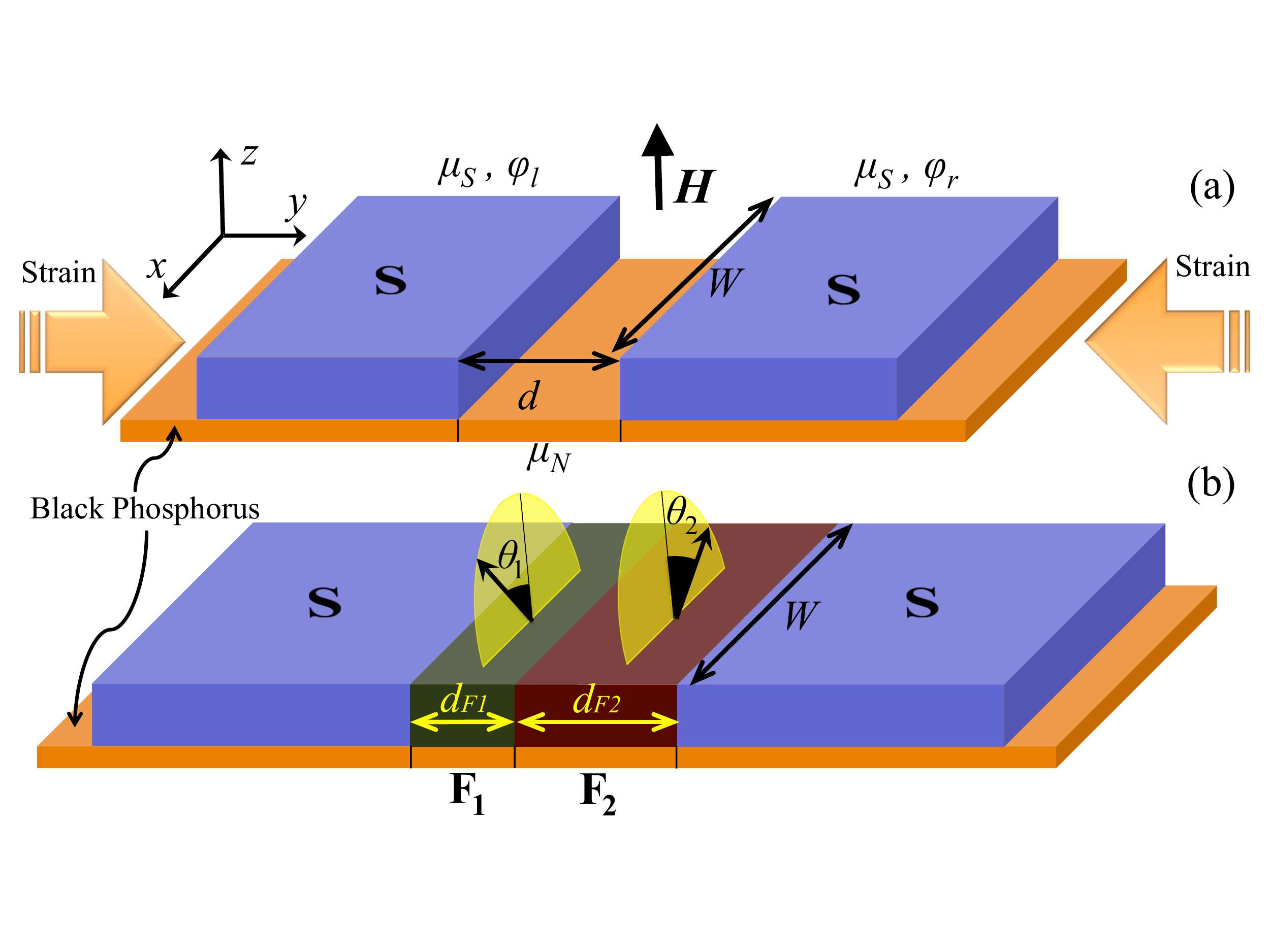}
\caption{\label{fig1} (Color online).
Schematic of the monolayer black phosphorus Josephson junctions. The two-dimensional system resides in the $xy$ plane, so that the interfaces are along the $x$ direction, and the $y$ axis is normal to the junctions. The uniaxial/biaxial strain is exerted into the plane of the black phosphorus sheet, and an externally applied magnetic field ${\bm H}$ is directed along $z$, perpendicular to the junction plane. The two $s$-wave superconducting electrodes (S), with different macroscopic phases $\varphi_{l,r}$, are proximity coupled to black phosphorus. The chemical potentials of superconducting parts and nonmagnetized, and magnetized regions are marked by $\mu_S$, $\mu_N$, and $\mu_{F1}, \mu_{F2}$. (a) The junction area is nonmagnetized BP of length $d$ and width $W$. (b) The junction area is magnetized through proximity coupling to ferromagnets ($\rm F_1$ and $\rm F_2$). The ferromagnets possess different lengths $d_{F1}\neq d_{F2}$ with identical widths $W$. The magnetization orientation in each F region is determined through its angle with respect to the $z$ axis, i.e., $\theta_{1,2}$. }
\end{SCfigure*}

The presence of defects and impurities when synthesizing BP sheets is inevitable and can influence the physical properties and characteristics of devices made from them. Therefore, the study of the influence of disorder at a microscopic level sheds light on the physical origins of experimental observations and facilitates the optimization of devices made of nonideal BP sheets\cite{R.Zhang,Koenraad,Doganov,Koenig,Han,Zou1,Liu2,qiu1}. In this paper, we establish physical models for hybrid structures made of superconducting and magnetized monolayer BP (from now on, by `black phosphorus' our mean is `monolayer black phosphorus'). We consider two-dimensional BP Josephson junctions with the possibility of inclusion of magnetism with arbitrary magnetization configurations, strain, and external magnetic field as depicted in Fig. \ref{fig1}. We study both ballistic and disordered devices containing nonmagnetic impurities. In ballistic systems, we develop a Bogoliubov-de Gennes microscopic theory and employ realistic band parameters for a BP sheet by means of density functional theory and symmetry calculations. We study the response of supercurrent to an external magnetic field and uniaxial/biaxial strain in a superconductor (S)-normal (N)-superconductor (S) BP junction where the superconductors are identical and possess different superconducting phases, creating a phase gradient $\varphi$ across the junction. Our results demonstrate that by manipulating the biaxial strain and Fermi level, the response of critical supercurrent to an external magnetic field deviates from the standard Fraunhofer interference pattern. Also, the critical current at current reversal points displays a finite nonvanishing supercurrent. In general, the current phase relation consists of multiple harmonics $\sin n\varphi$, $n= \pm 1, \pm 2, ...$. We plot the current phase relation around the current reversal points and show that the appearance of higher harmonics is responsible for the nonvanishing critical current. Additionally, we examine the influence of the magnetization misalignment angle on supercurrent flow in a superconductor (S)-ferromagnet ($\rm F_1$)-ferromagnet ($\rm F_2$)-superconductor (S) $\rm SF_1F_2S$ BP junction where F regions are of unequal thickness and have uniform magnetizations with misaligned orientations. Our investigations reveal a full spin switching of supercurrent upon increasing the magnetizations' relative misalignment angle $\theta$. For magnetizations smaller than $0.1$~eV, increasing the relative misalignment angle removes higher harmonics, while for stronger magnetizations ($\sim 0.2$~eV) it first induces higher harmonics close to $0$-$\pi$ crossovers at $\theta\sim\pi/2$ and then fully switches the supercurrent direction at antiparallel configuration, i.e., $\theta=\pi$. For disordered systems in the presence of magnetism and superconductivity, we formulate a quasiclassical Keldysh Green's function model. To this end, we utilize a low-energy model Hamiltonian and construct retarded, advanced, and Keldysh propagators. Applying the quasiclassical approximation, we derive the Eilenberger equation that is valid for ballistic and moderately disordered systems where the mean free time of particles $\tau$ is either infinite or sufficiently large. To further expand our theory, we consider the so-called diffusive regime, where the impurities scatter quasiparticles in all directions and the mean free time of particles goes to small values ($\tau\rightarrow 0$); expand the Green's function in terms of zeroth and first harmonics, and finally derive the Usadel equation. This approach was recently generalized for the surface channels of topological insulators (containing impurities and disorders) as well \cite{zyuzin,bobkova,alidoustphi0}. We derive charge current density in the diffusive regime and apply our theory to a two-dimensional SNS BP Josephson junction subject to an external magnetic field in the weak proximity limit, where the inducted superconducting gap into the normal region is less than $10\%$ of the gap deep inside the superconducting regions. Our results demonstrate that the delicate features explored in the ballistic regime, using the microscopic Bogoliubov-de Gennes theory, are washed out in the weak proximity limit of the diffusive regime, leading to the standard Fraunhofer response.

This paper is organized as follows. In Sec. \ref{sec:ballistic}, we study ballistic systems using a microscopic Bogoliubov-de Gennes formalism. Specifically, In Sec. \ref{subsec:mg_B_SNS}, we study the response of supercurrent to an external magnetic field, perpendicular to the junction plane, with the incorporation of biaxial and uniaxial strain and the variation of Fermi level. In Sec. \ref{subsec:SFFS_B}, we study the supercurrent in a $\rm SF_1F_2S$ BP Josephson junction and how magnetization rotation can alter the supercurrent. In Sec. \ref{sec:nonideal}, we present the generalized Eilenberger and Usadel theories for BP devices in the presence of superconductivity and magnetism. We apply this model to a SNS BP Josephson junction subject to an external magnetic field. Finally, we give concluding remarks in Sec. \ref{sec:conclusion}. 

\sidecaptionvpos{figure}{c}

\section{ballistic systems}\label{sec:ballistic}

The low-energy effective Hamiltonian of black phosphorus under strain $\varepsilon_{ii}$ and additionally subject to an external magnetic field with an associated vector potential $\text{A}_j$ can be expressed by\cite{Voon1,Voon2}
\begin{equation}\label{hamil1}
\begin{split}
&H= \int \frac{d\textbf{k}}{(2\pi)^2}\hat{\psi}^\dag({\textbf{k}})\Big\{ \big [u_0+\alpha_i\varepsilon_{ii} +(\eta_j+\beta_{ij}\varepsilon_{ii})(\text{k}_j-e\text{A}_j)^2\big]\tau_0 +\\&+\big[\delta_0+\mu_i\varepsilon_{ii} +(\gamma_j+\nu_{ij}\varepsilon_{ii})(\text{k}_j-e\text{A}_j)^2\big ]\tau_x -\chi_y (\text{k}_y-e\text{A}_y)\tau_y\Big\}\hat{\psi}({\textbf{k}}), 
\end{split}
\end{equation}
where the indices stand for coordinates, i.e., $i,j\equiv x,y$ , and summation over repeated indices is assumed. The band parameters calculated through density functional theory and symmetry computations are given in Table \ref{table}. When $\varepsilon_{ii}$ is negative (positive), the strain in that direction (the $x$ direction for $\varepsilon_{xx}$ and the $y$ direction for $\varepsilon_{yy}$) is of the compression (stretch) type. The matrices $\tau_i$ are the Pauli matrices in pseudo spin space, and an exchange field ${\bf h}$ invokes real-spin space indicated by $\text{h}_x\sigma_{x}+ \text{h}_y\sigma_{y}+ \text{h}_z\sigma_{z}$ in which $\sigma_i$ are the Pauli matrices in real-spin space. Therefore, in the presence of magnetism, the associated field operator is given by $\hat{\psi}^\dag(\textbf{k})=(\psi_{A\ua}^\dag, \psi_{A\da}^\dag,\psi_{B\ua}^\dag, \psi_{B\da}^\dag)$. Here the sublattices and real spins are labeled by $AB$ and $\ua\da$, respectively. 

We describe the superconductivity of a BP sheet through the BCS picture where particles with opposite spins are coupled by intervalley interactions. This type of pairing can be expressed by 
\begin{equation}
\Delta^{AB}_{\ua\da}\Big\langle\psi^\dag_{A\ua}\psi^\dag_{B\da}\Big\rangle+\text{H.c.},
\end{equation}
in which $\Delta^{AB}_{\ua\da}$ is the associated superconducting gap. Note that other pairing types can also be considered, as described in Ref. \onlinecite{BP_Majorana}, and these will influence the final results. Nonetheless, our investigations in Ref. \onlinecite{BP_Majorana} demonstrated that intravalley spin-singlet $s$-wave pairing has insubstantial influence. Other symmetry combinations such as $p$-wave, $d$-wave, $f$-wave, etc., may arise in BP under strain. The consequences of these combinations will be considered in a future work. There are some indications that a BP sheet under pressure and/or with electron doping can be driven into the superconducting phase \cite{Guo,Livas,Shirotani,Jun-JieZhang,YanqingFeng,Q.Huang,Xia,Wang1}, although still not conclusively \cite{R.Zhang,Yuan}. Nonetheless, we assume that singlet superconductivity can be extrinsically induced into a BP sheet by means of the proximity effect when BP is proximity coupled to an $s$-wave superconducting electrode as shown in Fig. \ref{fig1}. In this case, the superconducting BP sheet can be described by the following Bogoliubov-de Gennes Hamiltonian in the Nambu space:
\begin{equation}\label{bcs}
{\cal H}(\textbf{k})=\left( \begin{array}{cc}
H(\textbf{k})-\mu & \hat{\Delta}\\
\hat{\Delta}^\dag & -H^{\text{T}}(-\textbf{k})+\mu
\end{array}\right),
\end{equation}
where $\Delta$ is the proximity-induced superconducting gap and $\mu$ is the chemical potential. The associated $1\times 8$ vector field operator in $\textbf{k}$ space can now be expressed as $\check{\psi}_\text{BCS}^\dag(\textbf{k})=[\hat{\psi}^\dag(\textbf{k}),\hat{\psi}(\textbf{-k})]$. In the calculation of supercurrent below, the energies are given in units of $|\Delta|$.

\small
\begin {table}[b]
\caption {Band parameters of a monolayer black phosphorus subject to externally applied strain \cite{Voon1,Voon2}. } \label{table}
\begin{center}
\begin{tabular}{c*{4}{c}c}
\hline
\hline
$u_0$(eV) & $\delta_0$(eV) & $\alpha_x$(eV) & $\alpha_y$(eV) & $\mu_x$(eV)    \\
-0.42  & +0.76  & +3.15  & -0.58  & +2.65     \\
\hline
 $\mu_y$(eV) & $\eta_x$(eV$\cdot \textup{\AA}^2$)  & $\eta_y$(eV$\cdot \textup{\AA}^2$)  & $\gamma_x$(eV$\cdot \textup{\AA}^2$)  & $\gamma_y$(eV$\cdot \textup{\AA}^2$) \\
 +2.16  &  +0.58  & +1.01 & +3.93 & + 3.83  \\
\hline
$\beta_{xx}$(eV$\cdot \textup{\AA}^2$) & $\beta_{yx}$(eV$\cdot \textup{\AA}^2$) & $\beta_{xy}$(eV$\cdot \textup{\AA}^2$) & $\beta_{yy}$(eV$\cdot \textup{\AA}^2$)  \\
-3.48 & -0.57 & +0.80 & +2.39\\
\hline
$\nu_{xx}$(eV$\cdot \textup{\AA}^2$)  & $\nu_{yx}$(eV$\cdot \textup{\AA}^2$) & $\nu_{xy}$(eV$\cdot \textup{\AA}^2$) & $\nu_{yy}$(eV$\cdot \textup{\AA}^2$) & $\chi_y$(eV$\cdot \textup{\AA}$)  \\
-10.90 & -11.33 & -41.40 & -14.80 & +5.25\\
\hline
\hline
\end{tabular}
\end{center}
\end{table}
\normalsize

\subsection{SNS black phosphorus Josephson junction subject to external magnetic field}\label{subsec:mg_B_SNS}

\begin{figure*}
\includegraphics[ width=18.0cm,height=6.30cm]{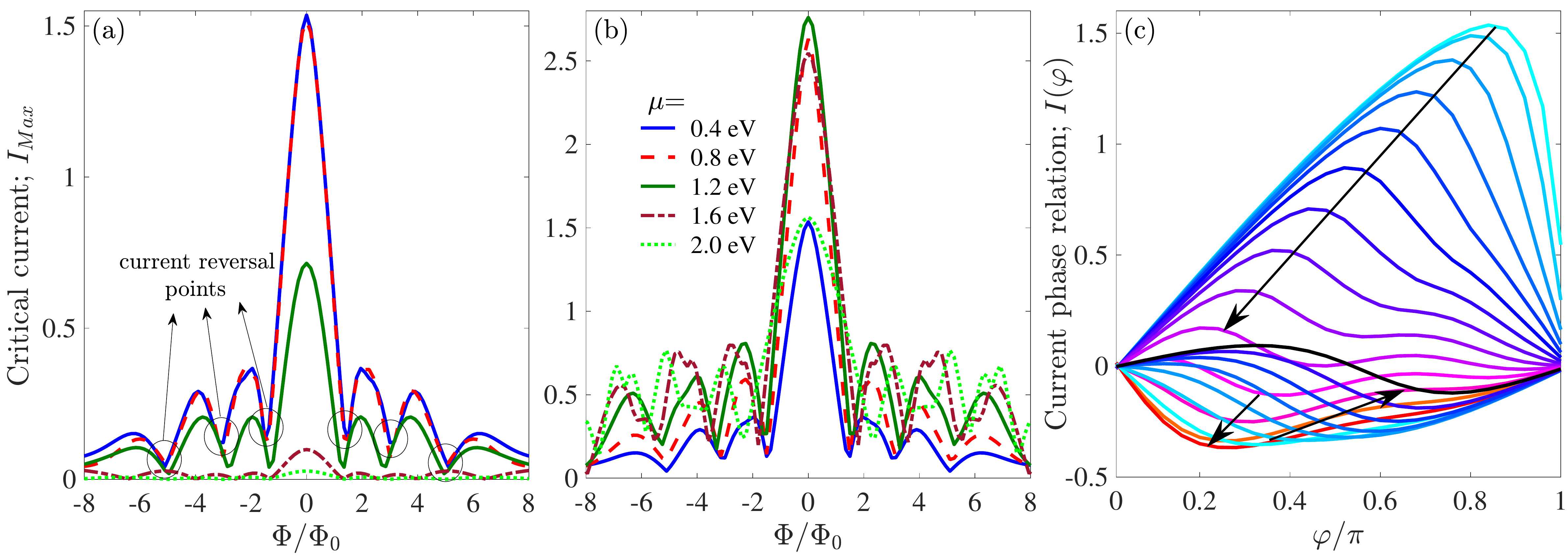}
\includegraphics[ width=18.0cm,height=6.30cm]{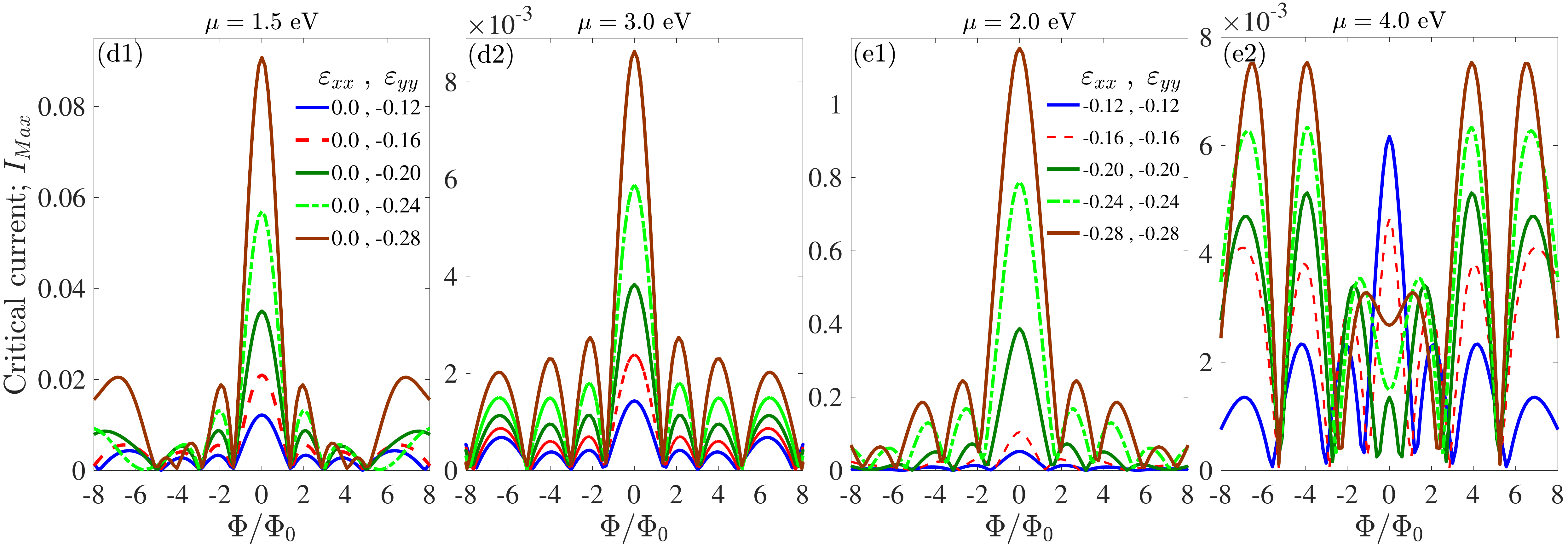}
\caption{\label{fig2} (Color online).
The response of critical current to an external magnetic field in a ballistic SNS BP Josephson junction. (a) The chemical potentials of both superconducting sides are varied, $\mu_S=\mu$, while the middle normal BP is undoped, $\mu_N=0$. The critical supercurrent is plotted as a function of magnetic flux passing through the junction area $\Phi$ (normalized by the magnetic flux quantum $\Phi_0=h/2e$) for different values of $\mu_S=\mu$ inside the S regions. The circles surround kinks where supercurrent reversals occur in cases with $\mu=0.4$~eV and $0.8$~eV. (b) The chemical potential throughout the junction is considered to be the same $\mu_S=\mu_N=\mu$ and the critical current is plotted for various values of $\mu$, $\mu=0.4, 0.8, 1.2, 1.6, 2.0$~eV. In (c), we show the current phase relation at a current reversal point that the external magnetic field induces. We consider a representative case, which is $\mu=0.4$~eV in panel (a). The arrows indicate the change in maximum supercurrent when varying the external magnetic field from $0$ to $\sim 3.0\Phi_0$. The compression in (a)-(c) is kept fixed at $\varepsilon_{xx}=0.0, \varepsilon_{yy}=-0.2$. In panels (d1)-(e2) we consider an undoped middle normal region $\mu_N=0$ and set differing scenarios for an applied strain. In (d1) and (d2) we set $\mu_S=\mu=1.5$ and $3.0$~eV in the superconducting regions and uniaxial stress $\varepsilon_{xx}=0.0, \varepsilon_{yy}=-0.12, -0.16, -2.0, -2.4, -2.8$. In (e1) and (e2), we have increased $\mu$ to $2.0, 4.0$~eV and set equal biaxial compression components in both the $x$ and $y$ directions: $\varepsilon_{xx}=\varepsilon_{yy}=-0.12, -0.16, -2.0, -2.4, -2.8$.  }
\end{figure*}

To begin, we consider a two-dimensional SNS BP Josephson junction as depicted in Fig. \ref{fig1}(a). The plane of BP is in the $xy$ plane, and the interfaces reside in the $x$ direction, so that the $y$ axis is perpendicular to the interfaces. The junction has a finite size in the $xy$ plane and has a length and width of $d$ and $W$, respectively. An external magnetic field $\bm H$ is exerted perpendicular to the junction plane and directed along the $z$ axis. To account for the external magnetic field, we consider a vector potential $\textbf{A}=(\text{A}_x,\text{A}_y,\text{A}_z)=(0,xH_z,0)$ so that $\textbf{A}$ satisfies the Lorentz condition $\bm\nabla \cdot\textbf{A}=0$ and produces the external field $ H_z=\bm\nabla \times \textbf{A}$ \cite{jap_ma}. We also consider the possibility of the inclusion of stress applied in the $x$ and $y$ directions. To evaluate the response of supercurrent to an external magnetic field, we calculate charge current density using the quantum definition, involving Hamiltonian (\ref{bcs}), i.e.,   
\begin{equation}\label{crntdif}
\begin{split}
\frac{\partial \rho}{\partial t}=\lim\limits_{{\bm r}\rightarrow {\bm r}'}\sum\limits_{\rho\tau\sigma\rho'\tau'\sigma'}\frac{1}{i\hbar}\Big[ \psi^\dag_{\rho\tau\sigma}({\bm r}'){\cal H}_{\rho\tau\sigma\rho'\tau'\sigma'}({\bm r})\psi_{\rho'\tau'\sigma'}({\bm r})\\-\psi^\dag_{\rho\tau\sigma}({\bm r}'){\cal H}_{\rho\tau\sigma\rho'\tau'\sigma'}^\dag({\bm r}')\psi_{\rho'\tau'\sigma'}({\bm r})\Big],
\end{split}
\end{equation}
where the time variation of charge density $\rho$ can be attributed to charge sources and sinks. Here ${\cal H}_{\rho\tau\sigma\rho'\tau'\sigma'}$ is the component form of Eq. (\ref{bcs}) with spin, valley, and particle-hole indices. Throughout the paper, we consider a steady-state regime where $\partial \rho/\partial t=0$. To compute the total current, we integrate the charge current density perpendicular to the interface $J_y$ over the $x$ direction, i.e., $I(\varphi)=\int_0^W dx J_y(x,y,\varphi)$. Note that the current is independent of the $y$-coordinate as required by charge conservation. We obtain appropriate spinors $\psi_{\rho\tau\sigma}$ within the normal BP region by matching wavefunctions $\hat{\psi}_l=\hat{\psi}_r$ and applying the continuity condition $\partial_\textbf{k} {\cal H}_l(\textbf{k})\hat{\psi}_l=\partial_\textbf{k} {\cal H}_r(\textbf{k})\hat{\psi}_r$, at the left and right superconductor interfaces. At the left boundary, ${\cal H}_l$ and ${\cal H}_r$ stand for the Hamiltonians of the superconducting and normal regions ($\hat{\psi}_l$ and $\hat{\psi}_r$ are their associated wavefunctions), respectively, whereas at the right boundary they stand for the Hamiltonians of the normal and superconducting regions, respectively. The resultant analytic expressions are very cumbersome, constituting $1\times 8$ spinors. We do not give the explicit expressions but employ them directly in evaluating observable quantities numerically. Supercurrent and its response to externally controllable agents such as magnetic field and stress are often measured in experiments. Figure \ref{fig2} illustrates the behavior of supercurrent passing through the device shown in Fig. \ref{fig1}(a). The junction length and width are fixed at $7$~nm and $20$~nm, respectively. On the one hand, the device experiences compression in the $y$ direction, i.e., $\varepsilon_{xx}=0, \varepsilon_{yy}=-0.2$. On the other hand, the junction is exposed to an external magnetic field where the magnetic flux penetrating the junction area is given by $\Phi=\pi dW{ H}_z$. Theoretically, it has been found that BP can support high strains as large as $40\%$ without any rupture or dislocations \cite{Carvalho,wei,Li1}. In Fig. \ref{fig2}(a), the middle segment of the SNS junction [Fig. \ref{fig1}(a)] is undoped, $\mu_N=0$, while the chemical potential in the superconducting segments is varied: $\mu_S=\mu=0.4, 0.8, 1.2, 1.6, 2.0$~eV. In Fig. \ref{fig2}(b) we set the chemical potential of the middle normal BP area equal to the superconducting parts, $\mu_N=\mu_S=\mu$. Increasing the chemical potential $\mu$ in the first case, the supercurrent is suppressed, while in the second case, the supercurrent first increases and then decreases. Our investigations of the density of states through the retarded Green's function, $N(\omega_n;\textbf{r})=-\pi^{-1}{\rm Im}\{{\rm Tr}[G^R(\omega_n;\textbf{r})]\}$, show that this behavior has a direct link to the nucleation of subgap bound states that the supercurrent flows through them (the so-called Andreev bound states). Increasing the number of Andreev states, and placing them closer to the superconducting gap edge increase the number of Cooper pairs that can pass across the junction from one superconductor to another under the applied superconducting phase gradient ($\varphi=\varphi_l-\varphi_r$). In both Figs. \ref{fig2}(a) and \ref{fig2}(b) we see that the critical supercurrent is nonvanishing even close to the supercurrent reversal points surrounded by circles. To illustrate the origin of nonvanishing critical current, we plot the charge current as a function of the superconducting phase difference between the two superconductors $\varphi$ for a gradually increasing $\Phi$ from $0$ to $\approx 3\Phi_0$ in Fig. \ref{fig2}(c). The supercurrent at zero flux $\Phi$ is proportional to $\sin\varphi$. By increasing $\Phi$, the supercurrent starts to deviate from the sinusoidal form. 
\sidecaptionvpos{figure}{c}
\begin{SCfigure*}{t!}
\includegraphics[ width=10.50cm,height=5.0cm]{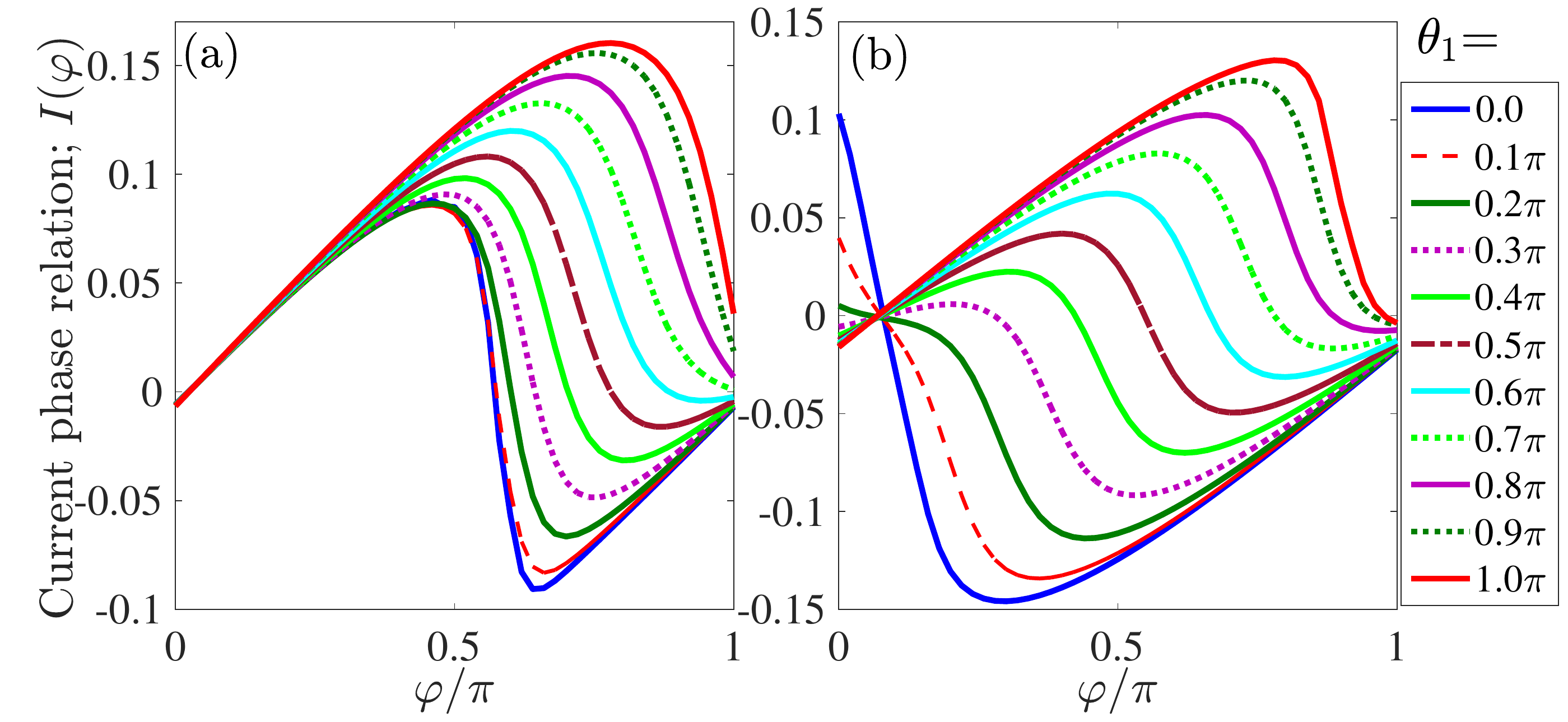}
\caption{\label{fig3} (Color online).
Current phase relation $I(\varphi)$ in a $\rm SF_1F_2S$ BP Josephson junction. The current is plotted with increasing magnetization misalignment angle $\theta_1=0.1\pi-1.0\pi$ with a step of $0.1\pi$. The magnetization in $\rm F_2$ is kept fixed along the $z$ direction, $\theta_2=0$. Hence, $\theta_1=\pi/2$ is equivalent to perpendicular magnetizations, while $\theta_1=\pi$ means antiparallel alignment of magnetizations. In panels (a) and (b) we set equal magnetization strengths in both $\rm F_1$ and $\rm F_2$, i.e., $|\textbf{h}_1|=|\textbf{h}_2|=\text{h}_0$ and change it as (a): $\text{h}_0=0.1$~eV, (b): $\text{h}_0=0.2$~eV.}
\end{SCfigure*}
The long downward arrow indicates the path that the maximum supercurrent traverses by increasing $\Phi$. Close to the first supercurrent reversal point in Fig. \ref{fig2}(a), the first harmonic $\sin\varphi $ is highly suppressed and higher harmonics $\sin 2\varphi, \sin 3\varphi, ... $ dominate. Note that because of the presence of these higher harmonics the supercurrent is always nonzero for all values of $\Phi$. A further increase in $\Phi$ induces a supercurrent reversal, as the small downward arrow indicates, and reverts the current phase relation to the one proportional to $\sin\varphi $, which is similar to the $\Phi=0$ case except with a minus sign. With a further increase in $\Phi$ the supercurrent mimics and repeats the behavior described above. Next, in Figs. \ref{fig2}(d1)-(e2) we examine the influence of uniaxial stress. In these panels we keep the middle BP undoped, $\mu_{F1}=\mu_{F2}=0$, and change $\mu_S=\mu$ in the superconducting regions to $1.5, 3.0, 2.0, 4.0$~eV from left to right. Figures \ref{fig2}(d1) and \ref{fig2}(d2) exhibit the effect of varying stress, $\varepsilon_{yy}=-0.12, -0.16, -0.20, -0.24, -0.28$, in the absence of any component in the $x$ direction; $\varepsilon_{xx}=0$. Increasing the compressive strain leads to a higher critical supercurrent. Interestingly, the contribution of higher harmonics to the supercurrent is now small and the critical supercurrent is vanishingly small at the supercurrent reversal points, which is the opposite of the cases shown in Fig. \ref{fig2}(b). In panels \ref{fig2}(e1) and \ref{fig2}(e2) we set an isotropic biaxial strain $\varepsilon_{xx}=\varepsilon_{yy}=-0.12, -0.16, -0.20, -0.24, -0.28$ with strengths identical to those in Figs. \ref{fig2}(d1) and \ref{fig2}(d2). As can be seen, the compressive strains $\varepsilon_{xx}$ and $\varepsilon_{yy}$ highly enhance the supercurrent when comparing Fig. \ref{fig2}(d1) to Fig.\ref{fig2}(e1). We also see that the contribution of higher harmonics close to the supercurrent reversal points is recovered, specifically for high strains $\varepsilon_{xx}=\varepsilon_{yy}= -0.24, -0.28$. The more prominent feature is shown in Fig. \ref{fig2}(e2). We see that not only do the higher harmonics prohibit nonzero supercurrent at the current reversal points, but also high strains $\varepsilon_{xx}=\varepsilon_{yy}= -0.20, -0.24, -0.28$ now result in non-Fraunhofer responses of supercurrent to the external magnetic field. At $\varepsilon_{xx}=\varepsilon_{yy}= -0.20$ the peak of the critical supercurrent at $\Phi=0$ is suppressed, and its amplitude is smaller than the second and third peaks. A further increase in compressive strains, i.e., $\varepsilon_{xx}=\varepsilon_{yy}= -0.24, -0.28$, causes a dip at zero external field, and the first peak of supercurrent appears near $\Phi\approx \Phi_0$. Note that the junction has a finite width of $20$~nm, and it results in nonideal Fraunhofer responses, as exhaustively studied in Ref. \onlinecite{jap_ma} for standard metallic systems.

\subsection{${\rm SF_1F_2S}$ black phosphorus Josephson junction}\label{subsec:SFFS_B}
  
In this section we expand our previous studies of ballistic systems to the $\rm SF_1F_2S$ devices depicted in Fig. \ref{fig1}(b). The magnetization and superconductivity can be induced in the BP sheet by means of the proximity effect. The two superconductors are coupled through a bilayer of uniformly magnetized ferromagnets, $\rm F_1F_2$, with different thicknesses $d_{F1}\neq d_{F2}$ and misaligned magnetization orientations $\theta_1\neq \theta_2$. The arrows on top of the F regions in Fig. \ref{fig1}(b) stand for the magnetization direction in each F layer that make angles $\theta_{1,2}$ with the $z$ axis perpendicular to the junction plane. Without loss of generality, we consider in-plane magnetizations, i.e., ${\mathbf h}=(\text{h}_x,0,\text{h}_z)$. The relative angle between the magnetization of the two F segments can be controlled in experiment by choosing different ferromagnetic materials for each F and applying an in-plane external magnetic field in the $x$ direction. It is known that various magnetic materials respond differently to an external magnetic field. The magnetization of strongly magnetized materials, e.g, $\rm Co$ or $\rm LCMO$ compounds, rotates harder and slower with respect to weakly magnetized materials, e.g., $\rm Py$ or $\rm NiFe$ in a given external magnetic field. An external magnetic field perpendicular to the junction plane can induce superconducting vortices that make analyses and the isolation of the pure effect of magnetization rotation inconclusive \cite{jap_ma,robin}. The in-plane external field in the $x$ direction, however, ensures that vortices are not generated, and therefore, if we choose $\rm F_1/F_2\equiv LCMO/Py$, the role of the external field is limited to only the induction of misalignment in the relative magnetization orientation of $\rm F_1/F_2$ \cite{robin}. Figure \ref{fig3} illustrates the behavior of the supercurrent phase relation $I(\varphi)$ when the magnetization misalignment angle increases. To obtain these results, we fix $d_{F1}=10$~nm and $d_{F2}=5$~nm and change the magnetization strength $|{\mathbf h}_1|=|{\mathbf h}_2|=0.1, 0.2$~eV in Figs. \ref{fig3}(a) and \ref{fig3}(b), respectively. We chose a representative strain set $\varepsilon_{xx}=0, \varepsilon_{yy}=-0.2$ and chemical potential $\mu_N=0, \mu_S=0.2$~eV. We set $\theta_2=0$ and vary $\theta_1$ from zero, equivalent to parallel alignment, to $\pi$, equivalent to antiparallel magnetization alignment, by a step of $0.1\pi$. In Fig. \ref{fig3}(a), when $\theta_1=0$, the supercurrent exhibits a sign change at a superconducting phase difference of $\varphi\approx 0.6\pi$. This current reversal occurs due to strong contributions of higher harmonics, i.e., $\sin 2\varphi, \sin 3\varphi, ...$. The magnetization misalignment weakens this contribution and removes the higher harmonics, so that at $\theta_1=0.6\pi$ the current reversal is fully removed, and when the magnetizations are antiparallel, we find the usual first harmonic $I(\varphi)\propto\sin \varphi$. The magnetization rotation from a parallel to antiparallel configuration causes a full switching of the supercurrent flow. This finding is suggestive of creating experimentally well controlled spin switching devices using the BP sheets. In Fig.~\ref{fig3}(b) when the magnetization strength is $0.2$~eV, we see that the weak phase shift at $\varphi=0$, also seen in Fig.~\ref{fig3}(a), is now pronounced. The magnetization misalignment removes it, induces a strong $\sin 2\varphi$ component at $\theta_1=0.5\pi$, and then at $\theta_1=\pi$ switches the supercurrent flow direction, including the phase shift at zero phase difference $\varphi=0$, which now reverts to a negative value. Other parameters such as junction thickness and temperature can also induce current reversals and higher harmonics in the current phase relationship.

\section{nonideal and disordered systems}\label{sec:nonideal}

To model nonmagnetic impurities, we formulate the quasiclassical model for a BP with the inclusion of superconductivity and magnetism. To this end, we express the Hamiltonian Eq. (\ref{hamil1})  by redefining parameters, which simplifies our subsequent notation, as follows
\begin{equation}\label{hamil2}
\begin{split}
H= \int \frac{d\textbf{k}}{(2\pi)^2}\hat{\psi}^\dag_{\textbf{k}}\Big\{ &\sum_{k=x,y}\eta_k(\text{k}_k-e\text{A}_k)^2\tau_0+\gamma_k(\text{k}_k-e\text{A}_k)^2\tau_x \\&-\chi_y (\text{k}_y-e\text{A}_y)\tau_y+\mu_0\tau_0+\mu_x\tau_x
\Big\}\hat{\psi}_{\textbf{k}}. 
\end{split}
\end{equation}
In the presence of spin and sublattices $A,B$, we define propagators \cite{antti}
\begin{subequations}\label{GF_comps}
\begin{eqnarray}
G_{\tau\sigma\tau'\sigma'}(t,t'; \mathbf{ r}, \mathbf{ r}') &=& - \Big\langle {\cal T}\Psi_{\tau\sigma} (t,\mathbf{ r}') \Psi_{\tau'\sigma'}^{\dag}(t',\mathbf{ r}')  \Big\rangle,~~~~\\
\bar{G}_{\tau\sigma\tau'\sigma'}(t,t'; \mathbf{ r}, \mathbf{ r}') &=& - \Big\langle {\cal T} \Psi^{\dag}_{\tau\sigma} (t,\mathbf{ r}) \Psi_{\tau'\sigma'}(t',\mathbf{ r}')  \Big\rangle,~~~~\\
F_{\tau\sigma\tau'\sigma'}(t,t'; \mathbf{ r}, \mathbf{ r}') &=& + \Big\langle {\cal T} \Psi_{\tau\sigma} (t,\mathbf{ r}) \Psi_{\tau'\sigma'}(t',\mathbf{ r}')  \Big\rangle,~~~~\\
F^{\dag}_{\tau\sigma\tau'\sigma'}(t,t'; \mathbf{ r}, \mathbf{ r}') &=& + \Big\langle {\cal T} \Psi^{\dag}_{\tau\sigma} (t,\mathbf{ r}) \Psi^{\dag}_{\tau'\sigma'}(t',\mathbf{ r}')  \Big\rangle,~~~~
\end{eqnarray}
\end{subequations}
where $\Psi_{\tau\sigma}$ are field operators, ${\cal T}$ is the time-ordering operator, and $t, t'$ are the imaginary times at $\mathbf{ r}, \mathbf{ r}'$ locations, respectively.
We consider the elastic scattering potential $V(\mathbf{ r})$ in a BP sheet by the self-energy term
\begin{equation}
\Sigma_{\text{imp}}(\mathbf{ r}-\mathbf{ r}')=\Big\langle V(\mathbf{ r})G(\mathbf{ r},\mathbf{ r}')V(\mathbf{ r}')\Big\rangle,
\end{equation}
where we average over the locations of impurities. To find the mean free time of particles in the disordered BP, we neglect anisotropic terms. Therefore, assuming isotropic scattering we find the mean free time as $\tau^{-1}=2\pi n_i N_0\int d\Theta_{{\bm n}_\text{F}}(2\pi)^{-1}|v(\Theta)|^2$, in which $v(\Theta)$ is the Fourier transform of the scattering potential that depends on the relative angle $\Theta$ between the particles' incidence direction and the particles' scattering direction, $n_i$ is the concentration of nonmagnetic impurities, and $N_0$ is the density of states per spin at the Fermi level of the system. In what follows, we assign the Pauli matrices $\sigma_{0,x,y,z}$ to real spin, $\tau_{0,x,y,z}$ to pseudo spin, and $\rho_{0,x,y,z}$ to the particle-hole in the presence of superconductivity. In the particle-hole space we find the following equation for the Green's function:
\begin{eqnarray}\nonumber \label{Green1}
\left( \begin{array}{cc}
-i\omega_n+\hat{H}(\mathbf{ r})& -\hat{\Delta}(\textbf{r}) \\
\hat{\Delta}^\dag(\textbf{r}) & i\omega_n +\tau_y\sigma_y\hat{H}^*(\mathbf{ r})\tau_y\sigma_y
\end{array} \right)\check{G}(\omega_n;\mathbf{ r},\mathbf{ r}')
\\
=\delta(\mathbf{ r}-\mathbf{ r}')+\frac{1}{2\pi N_0 \tau} \check{G}(\omega_n;\mathbf{ r},\mathbf{ r}) \check{G}(\omega_n;\mathbf{ r},\mathbf{ r}'),
\end{eqnarray}
in which $\omega_n=\pi (2n+1)k_BT$ is the Matsubara frequency, $n\in \mathbb{Z}$, $k_B$ is the Boltzmann constant, $T$ is temperature, and $\hat{\Delta}^\dag\rr$ is the Hermitian conjugation of $\hat{\Delta}\rr$: the proximity-induced superconducting gap.
The matrix form of the Green's function can be expressed by
\begin{equation}
\nonumber \check{G}(\omega_n;\mathbf{ r},\mathbf{ r}')=\left(  \begin{array}{cc}
-\hat{G}(\omega_n;\mathbf{ r},\mathbf{ r}') & -i\hat{F}(\omega_n;\mathbf{ r},\mathbf{ r}')\tau_y\sigma_y\\
-i\tau_y\sigma_y\hat{F}^\dag(\omega_n;\mathbf{ r},\mathbf{ r}') & \tau_y\sigma_y\hat{\bar{G}}(\omega_n;\mathbf{ r},\mathbf{ r}')\tau_y\sigma_y
\end{array}  \right).
\end{equation}
Here we have denoted $4\times 4$ matrices by using a hat symbol, $\hat{\square}$, and $8\times 8$ matrices by a check symbol, $\check{\square}$.  
We now subtract from Eq. (\ref{Green1}) its conjugate and Fourier transform with respect to the relative coordinates:
$\mathbf{ R}= (\mathbf{ r}+\mathbf{ r}')/2$ and $\delta \mathbf{ r} = \mathbf{ r}-\mathbf{ r}'$. In order to simplify our calculations, we assume that the Green's function is localized at the Fermi level and define the quasiclassical Green's function 
\begin{eqnarray}\label{G2}
\check{g}(\omega_n; \mathbf{ R}, \mathbf{ n}_\text{F}) =\frac{i}{\pi}\int d\xi_{ p} \check{G}(\omega_n; \mathbf{ R}, \mathbf{ p}),
\end{eqnarray}
in which $d\xi_p=v_Fdp$ and $v_F$ is the Fermi velocity. 
Incorporating these assumptions, we finally arrive at the Eilenberger equation \cite{eiln}:
\begin{gather}
p_\text{F}^k\Big\{{\cal J}_k, {\check{\tilde{ \nabla}}_k \check{g}\Big\}
 +\Big[\omega_n\rho_z-i\check{\Delta}-i\check{\text{M}}+i{\chi_y}p_\text{F}^y \tau_y+\frac{1}{2\tau} \langle \check{g} \rangle,\check{g}}\Big]=0,\nonumber\\
 {\cal J}_k=\eta_k+\gamma_k\tau_z,\label{eilenb}
 \\ \nonumber\check{\tilde{ \nabla}}_k \check{\text{X}}\equiv \check{\nabla}_k \check{\text{X}} - \Big[ie\text{A}_k\rho_z,\check{\text{X}}\Big],
\end{gather}
where the average over disorder is indicated by $\langle ... \rangle$ and $\check{\text{M}}$ stands for the magnetization. The momentum at the Fermi surface in the $k$ direction is shown by $p_F^k$. Here, we have assumed that the Fermi energy is large enough so that the Fermi wavelength, $\lambda_F$, is negligible compared to the spatial variation $\xi_{\bm r}$ of observable quantities, i.e., $\xi_{\bm r}\gg \lambda_F$.   

The Eilenberger equation can be further simplified in systems with a high density of impurities so that $ \tau^{-1} \gg |\omega_n| , |\Delta|$. 
In this case, the quasiparticles move diffusively with random directions and trajectories, which is the so-called diffusive regime \cite{usadel}. 
In the diffusive regime, we integrate the quasiclassical Green's function, Eq. (\ref{G2}), over all possible directions of the quasiparticles' momentum:
\begin{eqnarray}
\Big\langle \check{g}(\omega_n; \mathbf{ R}, \mathbf{ n}_\text{F}) \Big\rangle \equiv \int\frac{d\Omega_{\mathbf{ n}_\text{F}}}{2\pi}\check{g}(\omega_n; \mathbf{ R}, \mathbf{ n}_\text{F}), \;\mathbf{ n}_\text{F} =\frac{\mathbf{ p}_\text{F}}{|\mathbf{ p}_\text{F}|}. ~~~~~\;\;
\end{eqnarray}
In this regime, the Green's function can be expanded through the first two harmonics, $s$-wave and $p$-wave:
\begin{equation}\label{expansion}
\check{g} (\omega_n; \mathbf{ R},\mathbf{ n}_\text{F}) = \check{g}_s(\omega_n; \mathbf{ R}) + { n}_\text{F}^k \check{{ g}}_p^k (\omega_n; \mathbf{ R}),
\end{equation}
where the $s$-wave harmonic in the expansion is isotropic and much larger than the $p$-wave harmonic: $\check{g}_s \gg { n}_\text{F}^k \check{{ g}}_p^k$. By substituting this expanded Green's function into Eq. (\ref{eilenb}) and performing an integration over momentum directions we obtain
\begin{eqnarray}\label{Int1}
\check{ g}_p^k=-\tau p_\text{F}^k\check{g}_s\Big\{ {\cal J}_k,  \check{\tilde{ \nabla}}_k\check{g}_s \Big\}-\tau p_\text{F}^y\check{{g}}_s\Big[i\chi_y \tau_y , \check{g}_s \Big].
\end{eqnarray}
Next, we substitute Eq. (\ref{Int1}) into Eq. (\ref{eilenb}),  assume that $\nabla_k\gamma_k=\nabla_k\eta_k=\nabla_y\chi_y=\nabla_k\text{A}_k=0$, and find the generalized Usadel equation \cite{usadel}: 
\begin{equation}\label{Usadel}
\frac{p_\text{F}^k}{2}\Big\{ {\cal J}_k, \check{\tilde{ \nabla}}_k \check{{ g}}_p^k\Big\}+\frac{p_\text{F}^y}{2}\Big[  i\chi_y \tau_y, \check{{ g}}_p^k\Big]+\Big[\omega_n \rho_z-i\check{\Delta}-i\check{\text{M}},\check{g}_s \Big]=0.
\end{equation}

We next apply the quasiclassical model (\ref{Usadel}) to the SNS hybrid structure displayed in Fig. \ref{fig1}(a). We consider interfaces with a low transparency (the so-called tunneling limit) between the superconducting and normal BP regions. Therefore, the following boundary condition describes the coupling between the superconductor and the normal parts \cite{boundary_c1,boundary_c2}:
\begin{equation}\label{BC_sup}
\zeta { n}_k\check{{ g}}_p^k = \Big[\check{g}_s, \check{g}_{\text{SC}}\Big],
\end{equation}
in which $\zeta$ is the ratio between the resistance of the barrier region and the resistance in the normal region that controls the proximity effect at the boundaries, ${ n}_k $ is a unit vector perpendicular to the boundary, and $\check{g}_{\text{SC}}$ is the Green's function of the bulk superconducting segment. To study charge transport, we derive an expression for the charge supercurrent flow (due to the superconducting phase gradient across the device) in the $y$ direction [see Fig. \ref{fig1}(a)]. The quantum definition of current density, derived from Eq. (\ref{crntdif}), is expressed through the Hamiltonian, Eqs. (\ref{hamil2}). As stated in the previous section, we consider a steady state regime and therefore set $\partial \rho/\partial t=0$ in Eq. (\ref{crntdif}). After some calculations using Eq. (\ref{crntdif}), we finally arrive at the following expression for the current density in the diffusive regime: 
\begin{equation}\label{crnt_2}
{ J}_k({\bm r}) = \frac{ie\pi}{2} N_0\text{p}_\text{F}^kT \sum_n \mathrm{Tr} \Big[\rho_z \big(\eta_k+\gamma_k\tau_z\big)\check{{ g}}_p^k\Big].
\end{equation}
To obtain Eq. (\ref{crnt_2}) we have assumed a sufficiently small $\chi_y$ and neglected terms of the order of $\chi_y\text{p}_\text{F}^{-1}$ and once more assumed that $\xi_{\bm r}\gg\lambda_F$. We note that the formulated quasiclassical Eilenberger and Usadel formalisms can be extended to other two-dimensional materials. To the best of our knowledge, our work is the first that discusses the Eilenberger and Usadel approaches for BP and no counterpart of graphene type is yet available, which could be a potential subject for our future research \cite{zyuzin,bobkova,alidoustphi0}.  

\begin{figure}[t!]
\includegraphics[ width=7.50cm,height=5.90cm]{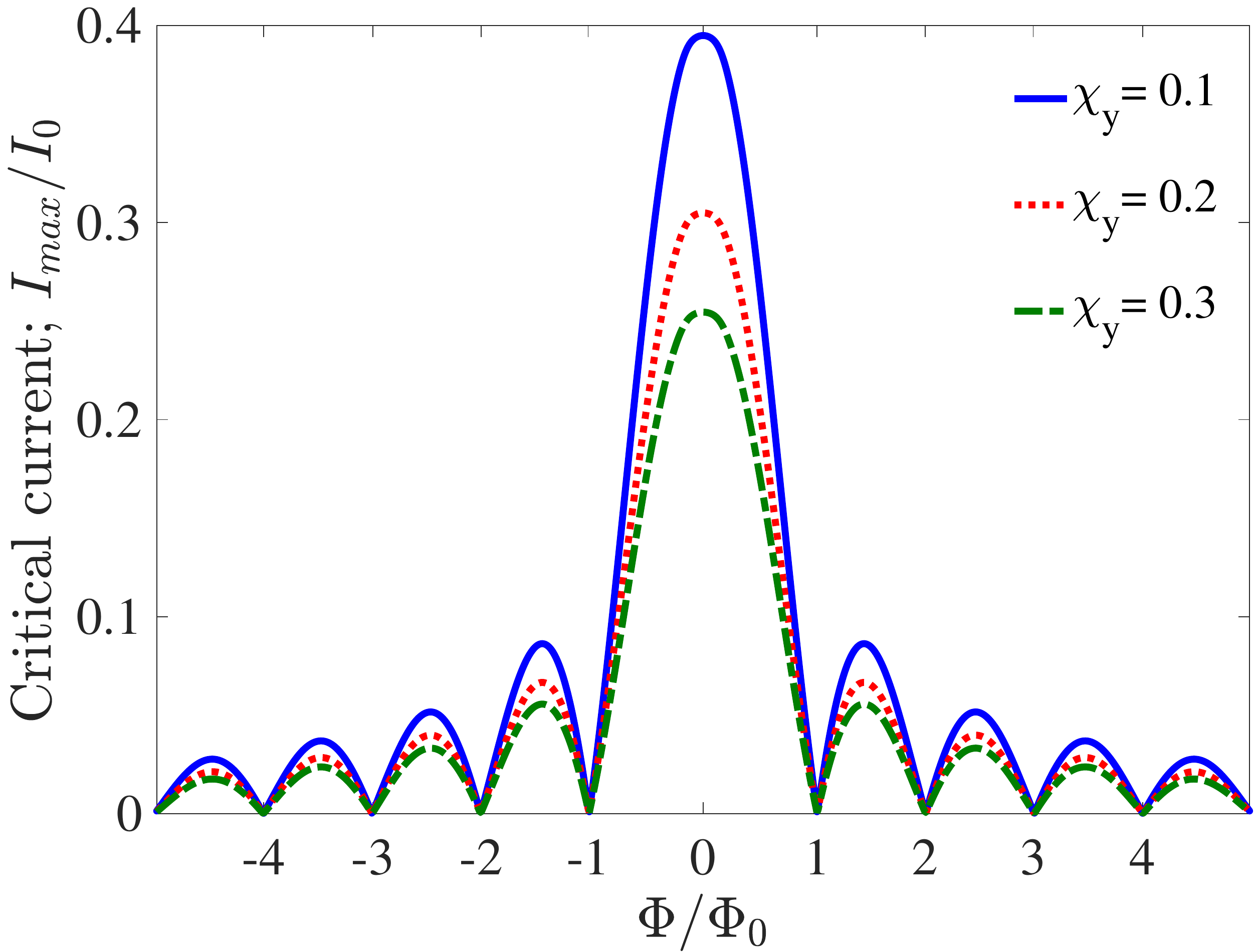}
\caption{\label{fig4} (Color online).
The critical current in a diffusive SNS BP Josephson junction as a function of external magnetic flux $\Phi$. Here we set $\eta=1$, $\gamma=0.5$ and vary $\chi_y=0.1, 0.2, 0.3$. }
\end{figure}

In order to find the charge current density flowing across the junction depicted in Fig. \ref{fig1}(a), we solve Eq. (\ref{Usadel}) together with proper boundary conditions, Eq. (\ref{BC_sup}), and substitute the resultant Green's function into Eq. (\ref{crnt_2}) for the diffusive regime. For the ballistic and/or moderately disordered systems, one solves Eq. (\ref{eilenb}) and uses the ballistic counterpart of the current density Eq. (\ref{crnt_2}) \cite{alidoustphi0}. In the diffusive regime, the Usadel equation (\ref{Usadel}) results in nonlinear boundary value differential equations that must be evaluated numerically \cite{vrtx_ali,jap_ma}. To further simplify our calculations, we have Taylor expanded the Green's function around its bulk solution $\check{g}_0$, i.e., $\check{g}_s\simeq \check{g}_0+\check{f}$. This limit is accessible in systems with weak proximity coupling between the superconductors and normal BP. Still, after performing these approximations, the resultant expressions are large coupled partial differential equations. Rather than presenting them explicitly, we evaluate them numerically \cite{jap_ma}. To calculate the supercurrent passing through the set up depicted in Fig. \ref{fig1}(a), we consider $\textbf{A}=(0,xH_z,0)$. The junction length and width are set to $d=\xi_S$ and $W=10\xi_S$, respectively, where $\xi_S$ is the superconducting coherence length. Figure \ref{fig4} shows the critical current as a function of applied magnetic flux $\Phi$. Because this simplified model is parametric and the quasiclassical approximations are applied, we have set representative values for parameters: $\eta=1, \gamma=0.5$ and $\chi_y=0.1, 0.2, 0.3$. The applied external magnetic field can induce a Zeeman field that can be large, depending on the $g$-factor of BP. The combination of Zeeman field and spin-orbit coupling can result in the generation of superconducting triplet correlations \cite{niu,Birge,Konschelle,Bergeret,ali_so1,ali_so2,H.Chakraborti}. Hence, we have considered a dominant spin-singlet superconductivity and added a small component of triplet pairings that can occur when making hybrid interfaces of superconducting BP subject to an external magnetic field. The critical current in the weak proximity limit of the diffusive regime shows the standard Fraunhofer patterns, and the variation of the parameters involved can simply change the overall amplitude of maximum current flow across the junction. As shown, in this weak proximity limit the delicate features explored in the ballistic regime (Sec. \ref{sec:ballistic}), such as nonzero supercurrent at the current reversal points and suppression of the central peak, vanish. Note that the results for the full proximity limit, however, can be closer to those explored in the ballistic limit (Sec. \ref{sec:ballistic}) and might retrieve the manifestation of the second harmonic discussed earlier \cite{vrtx_ali}. The methods developed in this work can be applied to the other limits (i.e., the full proximity limit, but accounting for impurities) and different geometries as well, as experimental structures emerge.

\section{conclusion}\label{sec:conclusion}
In summary, utilizing the low-energy effective Hamiltonian for black phosphorus, we formulated the Bogoliubov-de Gennes microscopic theory and quasiclassical model to describe ballistic and disordered black phosphorus samples in the presence of superconductivity and magnetism. We also incorporated an external magnetic field and biaxial/uniaxial stress in the models. Particularly, we studied the responses of supercurrent in a two-dimensional superconductor-normal-superconductor black phosphorus Josephson junction to an external magnetic field perpendicular to the junction plane. Our results demonstrate that by properly tuning the stress and Fermi energy, the critical supercurrent deviates significantly from the standard Fraunhofer interference pattern, so that the central peak is suppressed, while the amplitudes of the next peaks are larger than the central peak with nonzero current at supercurrent reversal points. We showed that the nonzero critical current at current reversal points is a direct consequence of the appearance of harmonics higher than the first usual harmonic in supercurrent around the current reversal points that prohibits fully vanishing current. Furthermore, we studied the influence of magnetization misalignment in a superconductor-ferromagnet-ferromagnet-superconductor junction where the ferromagnetic regions possess unequal thicknesses and magnetization orientations. Our results show a full supercurrent switching effect through increasing the magnetization misalignment angle. To complement our theory, we considered nonmagnetic impurities in the system. Employing the quasiclassical approximations, we derived Eilenberger and Usadel equations. The former is applicable to systems with a moderate density of impurities, while the latter describes systems containing a high density of impurities so that the quasiparticles have diffusive motions. We applied this model to a two-dimensional superconductor-normal-superconductor black phosphorus junction subject to an external magnetic field. Our investigation in the weak proximity limit of the diffusive regime showed the standard Fraunhofer response of supercurrent to an external magnetic field. Nevertheless, a full proximity limit can modify the results as this limit is closer to the ballistic regime. Our models and results provide a detailed explanation of the supercurrent behavior in BP Josephson junctions in the presence of impurities, external magnetic field, magnetic exchange field, and strain. Furthermore, our findings by the application of these models to specific configurations demonstrate that  both strain and the orientation of magnetic exchange fields offer effective tools to control the behavior of current in strain-effect BP transistors and highly sensitive BP-based devices such as superconducting quantum interference devices.      

\acknowledgments

M.A. is supported by Iran's National Elites Foundation (INEF). M.A. would like to thank A. Zyuzin and G. Sewell for useful discussions. Support from the Talent 1000 Foreign Expert program of China is acknowledged. Center for Nanostructured Graphene is supported by the Danish National Research Foundation (Project No. DNRF103).

\end{document}